\newcommand{\dint}{{\rm d}}
	\newcommand{\xpom}{x_{I\!\!P}}
	    \documentclass[%
	    reprint,
	    amsmath,amssymb,
	    aps,
	]{revtex4-1}
	    \usepackage{amsmath}
	    \usepackage{cleveref}
	    \usepackage{amsfonts}
	    \usepackage{graphicx}
	    \usepackage{dcolumn}
	    \usepackage{bm}
	
	    \usepackage{titlesec}
	  \usepackage{easyReview}
	    \titlespacing{\subsubsection}{2pt}{\parskip}{-\parskip}
	    
\begin{document}
	
	    	\title{Saturation and fluctuations in the proton wavefunction at large momentum transfers in exclusive diffraction at HERA}
	    	
	    	\author{Arjun Kumar}
	    	\email{arjun.kumar@physics.iitd.ac.in}
	
	   	\author{Tobias Toll}%
	   	\email{tobias.toll@physics.iitd.ac.in}

	    	\affiliation{%
	    		Department of Physics, Indian Institute of Technology Delhi, Hauz Khas, New Delhi 110016, India\\
	    	}
	
	    	\date{\today}
	
	    	\begin{abstract}
	We present a model of proton geometry where the number and size of gluon density hotspots in the proton's thickness function evolves with the resolution scale of the event given by the Mandelstam $t$ variable in exclusive diffractive $ep$ collisions. We use the impact-parameter dependent saturation dipole model bSat/IPSat, as well as its linearised (non-saturated) version bNonSat. In the latter the proton thickness has a clear interpretation as a thickness and in the former it is directly related to the saturation scale. The resulting phenomenological model for the splitting of hotspots, making full use of earlier experimental and phenomenological studies, is able to describe the entire incoherent $t$-spectrum for $|t|>1.1~$GeV$^2$ with a single phenomenological parameter. We use the previously suggested hotspot model as an initial condition for our evolution. The resulting model is a resolution scale-evolution in the same vein as a parton shower.The incoherent cross section is directly proportional to geometrical fluctuations in the proton's inital state. The hotspot evolution give rise to several kinds of event-by-event fluctuations such as in the hotspot number, width and normalization, and saturation scale fluctuations is a direct effect of these. A natural consequence of our resolution based evolution is that the hotspots obtain an effective repulsion. We use our hotspot evolution model to investigate saturation scale effects in the $t$-spectrum, and found that HERA data is not sensitive to this physics. 
	    	\end{abstract}
	    	\maketitle
	    	\section{\label{introduction}Introduction}
	    	Exclusive diffractive processes in Deeply Inelastic $ep$ Scattering (DIS) offer a unique window into the spatial structure of hadrons at small values of the Bjorken variable $x$. This has two reasons, firstly, small $x$ diffractive processes are (at leading order) directly proportional to the proton's \emph{longitudinal} gluon density \emph{squared}, which make these processes especially suitable for probing non-linear QCD effects such as saturation which can be described by the Color Glass Condensate (CGC) effective theory \cite{iancu2004color}. Secondly, these are the only processes in which the Mandelstam $t$-variable is experimentally accessible. At small $x$, DIS is best described by the dipole picture, in which the exchanged virtual photon splits into a quark-antiquark colour dipole. The square root of the $t$-variable is then the Fourier conjugate to the impact parameter $\textbf{b}$ between the dipole and the proton. Therefore, by measuring the $t$-spectrum one may get direct access (via a Fourier transform) to the proton's \emph{transverse} gluon structure in coordinate space. The impact parameter dependence has been implemented in many dipole models in the form of a two-dimensional thickness function $T(\vec{b})$ \cite{Kowalski:2003hm,Kowalski:2006hc,SCHLICHTING2014313,Mantysaari:2016ykx}. In the leading twist formulation of these models, $T(\vec{b})$ has a direct interpretation as the proton's thickness as a function of impact parameter. In the all twist saturation models, the thickness function can instead be interpreted in terms of the saturation scale, where $Q_S^2(x)\sim T(\vec{b})$.
	    
	    The exclusive diffractive events come in two classes: coherent and incoherent scattering, where in the former case the proton stays intact in the interaction, while in the latter the proton gets excited and subsequently dissociates. In the Good-Walker picture \cite{Good:1960ba}, the coherent cross section is proportional to the square of the first moment of the scattering amplitude, averaged over initial state configurations, while the incoherent cross section is given by the amplitude's initial state fluctuations. In the impact-dependent dipole model, the initial state fluctuations are geometrical fluctuations in the proton's transverse gluonic wave-function.

	In order to capture this, M\"antysaari and Schenke introduced a hotspot model \cite{Mantysaari:2016jaz, Mantysaari:2016ykx}. Here, the gluon wavefunction is concentrated in three hotspots and the position of these hotspots vary from event to event. The picture that emerges is that there are three fast valence partons at some large value of $x$ which act as sources for slow small $x$ partons where the latter become spatially correlated with the former. 
	 This geometric description of the proton proved effective in describing exclusive proton dissociative measurements of vector meson production at HERA for the first time in the dipole model. Here, the transverse size of the proton is probed at an areal resolution of $1/|t|$. As the hotspot model is a non-perturbative model of the gluon wave-function it was not expected to be valid for $|t|\gtrsim 1~$GeV$^2$, and it was indeed observed that the model underestimates the incoherent cross-section at larger momentum transfers. To address this discrepancy, additional fluctuations in the proton wave function are required at large momentum transfers. 
	 
	 	In \cite{Kumar:2021zbn} we showed that one may describe the entire measured $t$-spectrum ($|t|<30~$GeV$^2$) in exclusive diffraction by introducing a self-similar structure of gluon hotspots within hotspots. This introduced additional highly correlated size scales to the gluon structure for each such layer. While this approach is able to describe the entire $t$-spectrum well, and provides hints on what the transverse spatial gluon structure may look like, it lacks a mechanism for how such a structure emerges.
	 	
	In \cite{Demirci:2022wuy} the authors used a CGC approach in which pointlike colour charges, randomly distributed inside the geometry of the hotspots, were introduced, and they found a reasonable description of the whole $t$-spectrum. This calculation was performed in the non-relativistic and dilute limits so a quantitative match with measurements was not expected. Nevertheless, this approach may suggest three relevant scales of the gluon structure: the size of the proton, the size of the hotspots, and the pointlike size of the colour charges. However, this is not quite true as these pointlike charges give rise to hotspots within hotspots which correlate at the size of the saturation scale $Q_S$ which then is the actual third relevant scale, which will fluctuate from one point in transverse space to another, (almost) independently of $t$. It should be noted that the authors concluded that there is a need for additional fluctuations in the proton's wave function in order to describe the incoherent $t$-spectrum.
	
		Conceptually, there has been a recent suggested reinterpretation of the proton's wave-function.	As shown in e.g. \cite{Kharzeev:2017qzs,Hentschinski:2023izh}, the hadronic final state in $ep$ collisions at small $x$ exhibit maximum entropy, which, if one adopts the van Neumann interpretation of entropy, suggests that the small-$x$ partons in the proton are maximally entangled, and can thus be described by a common quantum mechanical wave function. 
	During an exclusive diffractive measurement, the transverse part of the gluon wave-function therefore collapses into an area of $\sim 1/|t|$. This will then be the relevant scale of fluctuations in that measurement.  
	An increased resolution therefore manifests as the hotspots obtaining a substructure of smaller hotspots at larger $|t|$, suggesting that hotspots in effect split into smaller hotspots with increased $|t|$. If the number of hotspots grow slower than their width decrease, or in other words, if the total cross-sectional area of the proton decreases with $|t|$, then at least some hotspots get hotter, which manifests as larger saturation scale fluctuations. 
	
	The aim of this paper is to formulate a hotspot evolution in a numerical model which is consistent with all physics considerations discussed above: The self-similarity of the proton's gluon content at different resolution scales, the need for more sources of fluctuations than what comes out of a direct CGC approach, and the maximal entanglement of the small $x$ gluons. Our model should describe the entire $t$-spectrum in exclusive $J/\psi$ production at HERA, treating the original hotspot model as an initial state of the evolution. As will be seen, this phenomenological approach only needs one parameter in order to describe all the data beyond the applicability of the hotspot model. We also investigate which saturation effects may be revealed from these measurements based on our model. 
	
	This paper is organised as follows. In the next section we give a brief description of the dipole models and the hotspot model. In section \ref{sec:evolution} we describe our hotspot evolution model and in section \ref{sec:results} we show and discuss the resulting cross sections as well as give an extrapolation of the model to heavy nuclei, before we end with our conclusions and outlook.
	\section{ The Colour Dipole Models}
		\begin{figure}
		\centering
		\includegraphics[width=0.7\linewidth]{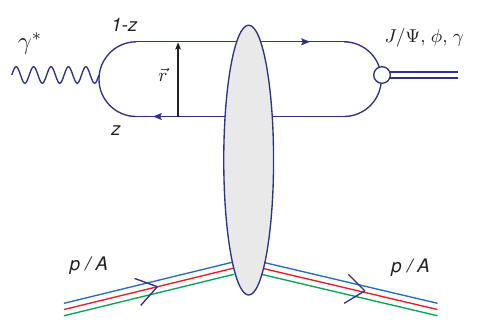} 
		\caption{$\gamma^*p$ scattering cross section in the dipole model \cite{PhysRevD.105.114045}}
		\label{dipole}
	\end{figure}
	In the dipole picture, the diffractive vector meson production's scattering amplitude results from a convolution of three subprocesses, illustrated in Fig.~\ref{dipole}. Initially, the virtual photon undergoes a splitting into a quark-antiquark pair forming a color dipole. Subsequently, this dipole interacts with the proton through one or multiple colorless two-gluon exchanges. Finally, the dipole recombines, forming a vector meson. The corresponding amplitude is expressed as follows:
	\begin{eqnarray}	\label{eq:amplitude}
	\mathcal{A}_{T,L}^{\gamma^* p \rightarrow J/\Psi p} (x_{\mathbb{P}},Q^2,{\bf \Delta})
	&=&i\int \dint^2 {\bf r}\int \dint^2{\bf b}\int \frac{\dint z}{4 \pi}  \nonumber \\
	&\times& (\Psi^*\Psi_V)_{T,L}(Q^2,{\bf r},z) \\  
	&\times& e^{-i[\bf{b}-(\frac{1}{2}-z){\bf r}].{\bf \Delta}} \frac{\dint\sigma _{q\bar{q}}}{\dint^2{\bf b}}({\bf b},{\bf r},x_\mathbb{P})\nonumber
	\end{eqnarray}
	where $T$ and $L$ represent the transverse and longitudinal polarizations of the virtual photon respectively, \textbf{r} denotes the transverse size of the dipole, and \textbf{b} is the impact parameter of the dipole relative to the proton's center. The variable $z$ is the light-cone momentum fraction of the photon taken by the quark, and ${\bf \Delta} \equiv \sqrt{-t}$ is the transverse part of the four-momentum difference between the outgoing and incoming proton. The term $(\Psi^* \Psi_V)$ is the wave-function overlap between the virtual photon and the produced vector meson. 
	The dipole cross section, denoted as $\frac{d\sigma _{q\bar{q}}}{d^2\textbf{b}}(\textbf{b},\textbf{r},x_\mathbb{P})$, characterizes the strong interaction between the dipole and the proton and is discussed below. It is twice the scattering amplitude $\mathcal{N}(\textbf{b},\textbf{r},x_\mathbb{P})$. The vector-meson wave functions are typically modelled, and for $J/\psi$, a boosted Gaussian wave function is employed with parameter values from \cite{Mantysaari:2018nng}. Note that the amplitude is a Fourier transform from the transverse coordinate of the quark in the dipole to the transverse momentum transfer of the proton. Consequently, the resulting cross section carries information about the spatial structure of the proton and its quantum fluctuations. 
	
	We consider here the saturated (bSat, also called IPSat) and non-saturated (bNonSat) versions of the dipole cross-section. The bSat model dipole cross section is given by 
	\cite{Kowalski:2003hm}:
	\begin{eqnarray}
	\frac{\dint\sigma _{q\bar{q}}}{\dint^2\textbf{b}}(\textbf{b},\textbf{r},x_\mathbb{P})=
	2\big[1-\text{exp}\big(-F(x_\mathbb{P} ,\textbf{r}^2)T_p(\textbf{b})\big)\big]
	\end{eqnarray}
	with
	\begin{eqnarray}
	F(x_\mathbb{P} ,\textbf{r}^2)=\frac{\pi^2}{2N_C} \textbf{r}^2 \alpha_s(\mu^2) x_\mathbb{P} g(x_\mathbb{P},\mu^2),
	\end{eqnarray}
	Due to the exponential form, this model saturates the cross-section for large gluon density $xg(x, \mu^2)$, large dipole sizes $r$, or for large values of the thickness function $T_p(\textbf{b})$. This defines the saturation scale $Q_S(r, x, \textbf{b})=2/r_S(r, x, \textbf{b})$ where $r_S$ is the value of $r$ which solves $F(\xpom, \textbf{r}^2)T(\textbf{b})=1/2$. Note that $Q_S^2\sim T_p(b)$, i.e. that the saturation scale is (nearly) proportional to the thickness function of the proton. 
	The scale at which the strong coupling $\alpha_s$ and gluon density is evaluated at is $\mu^2 = \mu_0^2 +\frac{C}{r^2}$ and the gluon density at the initial scale $\mu_0$ is parametrised as:
	\begin{eqnarray*}
		x g(x,\mu_0^2)= A_g x^{-\lambda_g}(1-x)^{6}
	\end{eqnarray*}
where the parameters $A_g$, $\lambda_g$, $C$, and $m_f$ are determined through fits to inclusive DIS reduced cross section measurements. We use the fit results from \cite{Sambasivam:2019gdd}. The transverse profile of the proton is usually assumed to be Gaussian:
	\begin{eqnarray}
	T_p(\textbf{b}) = \frac{1}{2 \pi B_G}\exp\bigg(-\frac{\textbf{b}^2}{2B_G}\bigg)
	\label{eq:profile}
	\end{eqnarray}
	and the parameter $B_G$ is constrained through a fit to the $t$-dependence of the exclusive J/$\psi$ production at HERA \cite{Kowalski:2006hc}, and is found to be $B_G=4.25~\pm~0.4~$GeV$^{-2}$. Note that the central value of the parameter $B_G$ is changed slightly from the previous value of $4~$GeV$^{-2}$ to offset the effect of the new phase factor in eq.(\ref{eq:amplitude}). 
	
	The linearised dipole cross section, the bNonSat model, is given by: 
	\begin{equation}
	\frac{\dint\sigma _{q\bar{q}}}{\dint^2\textbf{b}}(\textbf{b},\textbf{r},x_\mathbb{P})=\frac{\pi^2}{N_C}\textbf{r}^2\alpha_s(\mu^2) x_\mathbb{P} g(x_\mathbb{P},\mu^2)  T_p(\textbf{b})
	\end{equation}
	which does not include any saturation effects and gives only the leading twist contribution. This dipole cross-section corresponds to a single two-gluon exchange. Currently, both models effectively describe HERA $F_2$ and exclusive data \cite{Mantysaari:2018nng,Rezaeian:2012ji}, while the studies of exclusive vector meson production in ultra-peripheral collisions of lead nuclei at the LHC exhibit a clear preference for the bSat model \cite{Sambasivam:2019gdd}.
	
	The exclusive cross section receives large corrections (discussed in detail in \cite{Kowalski:2006hc, Mantysaari:2016ykx}). Firstly, the scattering amplitude in eq.~\eqref{eq:amplitude} is approximated to be purely imaginary. However, the real part of the amplitude is taken into account  by multiplying the cross section by a factor of (1+$\beta^2$) with $\beta = \tan\big(\lambda \pi/{2})$, and $\lambda = \partial \log (\mathcal{A}_{T,L}^{\gamma^*p\rightarrow Vp})/\partial \log(1/x)$. Secondly, to take into account that the two gluons may have different momentum fractions, a skewedness correction to the amplitude is introduced \cite{Shuvaev:1999ce}, by a factor $R_g(\lambda)= 2^{2 \lambda +3}/\sqrt{\pi} \cdot\Gamma (\lambda_g + 5/2)/\Gamma (\lambda_g+4)$ with $\lambda_g= \partial \log (xg(x))/\partial \log(1/x)$. In our model, we calculate both these corrections using a spherical proton. 	
	
	The elastic diffractive cross section for a smooth spherical proton without any fluctuations in its wave function (for the thickness function provided in eq.(\ref{eq:profile})), is given by:
	\begin{equation}
	\frac{d \sigma^{\gamma^* p \rightarrow J/\Psi p}}{dt} = \frac{1}{16 \pi} \big| \mathcal{A}^{\gamma^* p \rightarrow J/\Psi p} \big|^2
	\end{equation}
	When fluctuations are included in the wave function of the target, as discussed in \cite{Mantysaari:2016ykx,Mantysaari:2016jaz,Kumar:2022aly,Kumar:2021zbn} for the proton, we employ the Good-Walker formalism \cite{Good:1960ba}. In the Good-Walker picture, the coherent cross section is proportional to the first moment of the amplitude and probes the average shape, while the total cross section is sensitive to the second moment. The incoherent cross section is the difference between the second moment and first moment squared, which for Gaussian distributions is its variance and as such probes the fluctuations in the target wave function. Thus, for an event-by-event variation $\Omega$, we have:
	\begin{eqnarray}
	\frac{{\rm d} \sigma_{\rm coherent}}{{\rm d}t} &=& \frac{1}{16 \pi} \big| \left<\mathcal{A}(x_{\mathbb{P}},Q^2,\textbf{$\Delta$})\right>_\Omega\big|^2 \nonumber \\
	\frac{{\rm d} \sigma_{\rm incoherent}}{{\rm d}t} &=& \frac{1}{16 \pi}\bigg(\big< \big| \mathcal{A}(x_{\mathbb{P}},Q^2,\textbf{$\Delta$})\big|^2\big>_\Omega \\\nonumber&~& ~~- \big| \big<\mathcal{A}(x_{\mathbb{P}},Q^2,\textbf{$\Delta$})\big>_\Omega\big|^2\bigg)
	\end{eqnarray}
	This approach has previously been employed in the context of heavy nuclei, where nucleons are considered Gaussian hotspots and the transverse positions of nucleons are distributed based on the Woods-Saxon potential in a Glauber-like description. \cite{Toll:2012mb,Toll:2013gda,Mantysaari:2017dwh,Mantysaari:2022sux}.
	
		\subsection*{The Dipole Model in Heavy Nuclei}
In order to extrapolate the dipole models to heavy nuclei we use the independent scattering model, such that the scattering amplitude for the nucleus is given by \cite{Toll:2012mb}:
	\begin{eqnarray}
	1-\mathcal{N}^{({\rm A})}(\textbf{b},\textbf{r},x_\mathbb{P})=
	\prod_{i=1}^A \left(1-\mathcal{N}^{(p)}(\textbf{b}-\textbf{b}_i,\textbf{r},x_\mathbb{P})\right)
	\end{eqnarray}
where the transverse positions of the nuclei, $\textbf{b}_i$, are sampled from a Woods-Saxon distribution. This yields the following dipole cross section:
\begin{eqnarray}
	\frac12\frac{\dint\sigma _{q\bar{q}}^{({\rm A})}}{\dint^2\textbf{b}}(\textbf{b},\textbf{r},x_\mathbb{P})=
	1-\text{exp}\big(-F(x_\mathbb{P} ,\textbf{r}^2)\sum_{i=1}^A T_p(\textbf{b}-\textbf{b}_i)\big) \nonumber
\end{eqnarray}
	Thus, the nuclear thickness function can be written as a sum of the individual nucleons' thickness functions:
	\begin{eqnarray}
		T_A(\textbf{b})=\frac{1}{A}\sum_{i=1}^A T_p(\textbf{b}-\textbf{b}_i)
	\end{eqnarray}
	where we have normalised the nuclear thickness function to unity.
	
	\subsection*{The Hotspot model}
	In the hotspot model, the proton is characterized by a lumpy configuration consisting of hotspots of gluonic density at small-$x$. In a simplified representation, one may envision three hotspots, hypothesized to emerge from the evolution of small-$x$ gluons around three valence quarks at large-$x$ acting as sources. The gluon density within these hotspots is assumed to be isotropic and smooth, implying a single relevant scale corresponding to the hotspot width for incoherent diffraction. Again deploying the independent scattering approximation, the transverse profile of the proton in the hotspot model is given by \cite{Mantysaari:2016ykx}:
	\begin{eqnarray}
	T_p(\textbf{b}) = \frac{1}{N_{hs}}\sum_{i=1}^{N_{hs}}T_{hs}(\textbf{b-b$_i$}),
	\end{eqnarray}
	with
	\begin{equation}
	T_{hs}(\textbf{b}) = \frac{1}{2 \pi B_{hs}}\exp\big[-\frac{\textbf{b}^2}{2B_{hs}}\big]
	\label{eq:hsprofile}
	\end{equation}
	Here, $N_{hs}$ is set to 3, and $\textbf{b}_i$ represents the locations of hotspots, sampled from a normalised Gaussian of width $B_{qc}$. The width of the hotspots is $B_{hs}$. The parameters $B_{qc}$ and $B_{hs}$ govern the degree of fluctuations in the proton wave function, and their values are constrained by both coherent and incoherent data. It should be noted that since the saturation scale $Q^2_S\sim T(\textbf{b})$, the saturation scale is larger in the hotspots, and that for some configurations of hotspots we may get a rather large saturation scale compared to what is possible for a spherical proton. 
	
	\begin{figure*}
		\centering
		\includegraphics[width=0.85\linewidth]{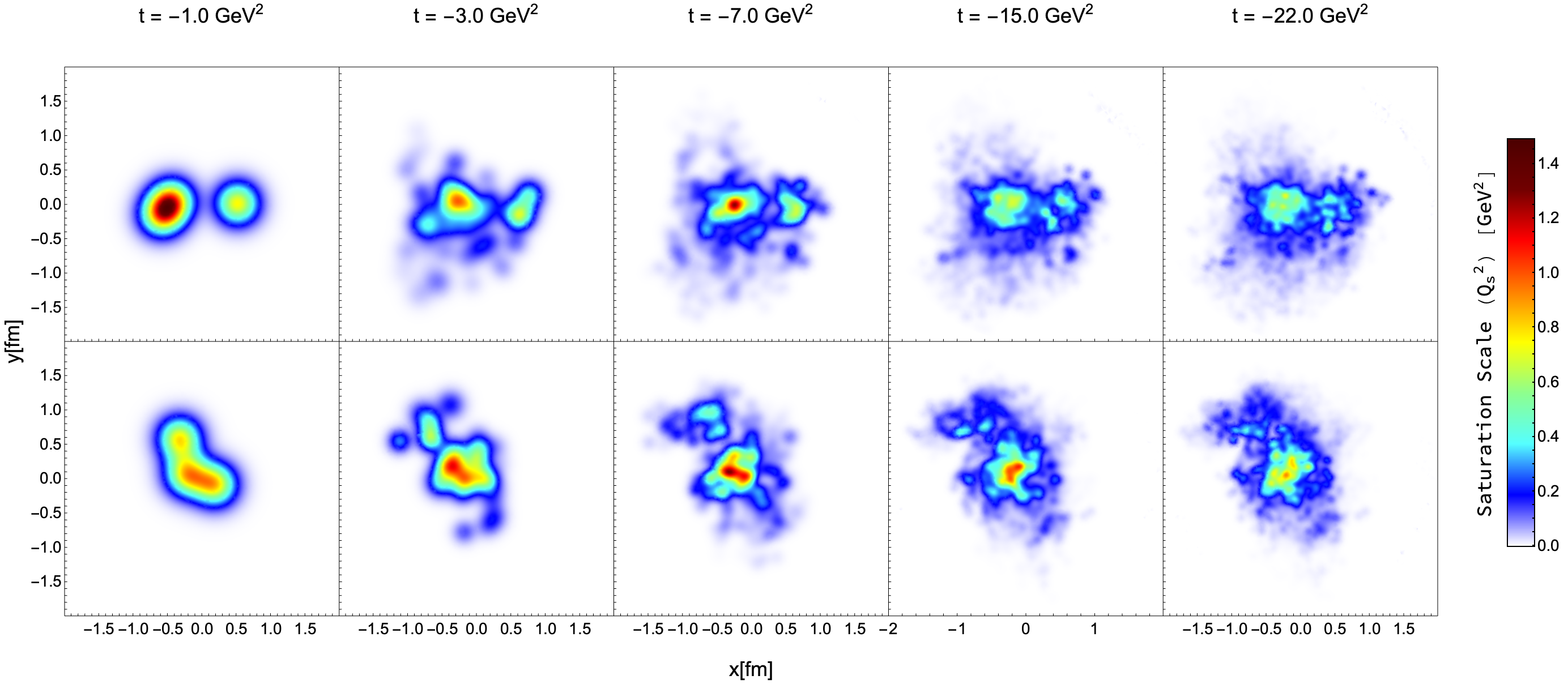}
		\caption{The resulting saturation scale $Q_S$ as a function of transverse coordinates at different values of $t$ for two initial state hotspot distributions of the proton. Here, $\xpom=10^{-4}$.}
		\label{fig1}
	\end{figure*}
	
	\section{The Hotspot Evolution model}
	\label{sec:evolution}
	A measurement at a certain Mandelstam variable $t$ at small $x$ constitute a measurement of the combined gluonic wave function at an areal resolution $1/|t|$. Therefore, the measurement causes the gluon wave function to collapse into an area $B\sim 1/|t|$. At larger $|t|$ we would therefore measure a smaller area of the gluon wave-function. This manifests as a splitting of a hotspot existing at one $|t|$ into two hotspots at a larger $|t|$. 
	In order to capture this resolution picture of the proton's gluonic substructure, we suggest the following hotspot evolution process:
	\begin{itemize}
		\item Let the original non-perturbative hotspot model act as an initial state of the evolution at $|t|\leq |t_0|$.
		\item Evolve the initial state by probabilistically splitting the hotspots based on the resolution for $|t|>|t_0|$.
	\end{itemize}
	In order to evolve the initial state by splitting of the hotspots one needs to define a probability density distribution, a so called splitting function. We know from the self-similarity found in \cite{Kumar:2021zbn} that the hotspot structure should scale with $\ln |t|$. This suggests a simple model where $\dint\mathcal{P}_{\rm split}(t)=\alpha/|t| \dint t$. To find the probability of a hotspot that exists at a scale $t_1$ to split at scale $t_2$, one needs to multiply this splitting function at $t_2$ with the probability density for  no splitting occurring in the interval $t_1\leq t\leq t_2$, where the latter probability is given by:
	\begin{eqnarray}
		\frac{\text{d}{\mathcal{P}}_{\rm no-split}}{\text{d}t }=&\text{exp}\bigg(-\int_{t_0}^t \text{d}t' \frac{\text{d}{\mathcal{P}}_{\rm split}}{\text{d}t' }\bigg)\nonumber
	\end{eqnarray}
	This simplest model gives non-physical large tails in the hotspot distribution as some hotspots become very long-lived. Instead we are using the following hotspot evolution model:
	\begin{align}
	\frac{\text{d}{\mathcal{P}}_{\rm split}}{\text{d}t } =& \frac{\alpha}{|t|} \frac{t-t_0}{t}\nonumber\\ 
	\frac{\text{d}{\mathcal{P}}}{\text{d}t } =& \frac{\alpha}{t}\frac{t-t_0}{t} \text{exp}\bigg[-\alpha \bigg(\frac{t_0}{t} -\text{ln}\frac{t_0}{t}-1\bigg)\bigg]
	\label{eq:TM}
	\end{align}
	\noindent where $t_0$ is the initial scale and $\alpha$ is the evolution parameter which is determined by fits to the measured $t$-spectra.
	We see that our model gives a maximum probability density hump which is somewhat displaced from $t=t_0$. This means that the hotspots widths fluctuate around the resolution scale $1/|t|$. The model has the desired asymptotic behaviour for large $|t|$ where $\dint\mathcal{P}_{\rm split}/\dint t\propto 1/|t|$. 
	The  algorithm for the evolution of the initial state is as follows:
	\begin{itemize}
		\item Offspring hotspots $i$ and $j$ are created at distance $d_{ij} = |\textbf{b}_i-\textbf{b}_j|$ sampled from the parent hotspot with widths $B_{i,j} = \frac{1}{|t|}~\text{GeV}^{-2}$, where the $t$ value is sampled from eq.\eqref{eq:TM}.
		\item We impose probe and geometry resolution criteria (which coincide in the case of Gaussian hotspots), by demanding that $d_{ij} >2 \sqrt{B_{i,j}}$. If this is not the case, the two offspring hotspots cannot be resolved by the probe and therefore, the splitting is rejected. This results in an effective hotspot repulsion.  
	\end{itemize}
	A hotspot repulsion, or a minimum possible core distance in three dimensions between hotspots, $r_c$, was introduced ad hoc in \cite{Albacete:2016pmp, Albacete:2017ajt} in order to explain the hollowness effect in the proton, which in turn is used to explain the anti-correlation between $v_2$ and $v_3$ flow harmonics at the highest multiplicities in $pp$ collisions. For three hotspots, they found that the values of $r_c$ lie in the range $r_c/\sqrt{B_{i, j}}\in [0.7,1.7]$ which is slightly smaller than the transverse distance in our model. In a Bayesian analysis of the hotspot parameters compared to exclusive diffractive HERA data \cite{Mantysaari:2022ffw}, it was however found that the description of the data is not sensitive to the value of $r_c$ in the range where $r_c/\sqrt{B_{i, j}}\in [0.0, 7.9]$. It should be noted that this repulsion emerges naturally from our approach, as a resolution effect. 
	
	The total thickness function of the proton is calculated as:
	\begin{eqnarray}
	 T_p(\textbf{b}) = \frac{1}{N_{hs}}\sum_{i=1}^{N_{hs}}T_i(\textbf{b-b$_i$})
	\end{eqnarray}
	where $N_{hs}$ is the total number of resolved hotspots at that scale. We make sure that the total normalisation of this thickness function is conserved throughout the evolution. 
	The evolution of the proton profile for two distinct events are shown in Fig.~\ref{fig1}, in which we plot the resulting saturation scale $Q_S(x, \textbf{b})$ at different resolutions $|t|$. At large momentum transfers, we observe the emergence of additional features indicative of small-scale fluctuations. It is noteworthy that our representation of the proton at very high momentum transfers bears resemblance to the color glass condensate description of the proton initial state where its structure is characterized by point-like color charges superimposed on the geometric hotspot structure \cite{Mantysaari:2020axf}. To compute cross sections, we generate 800 proton configurations from our model. 
	
	This model contains several different kinds of event-by-event fluctuations, namely hotspot width, hotspot number, and normalization fluctuations. Since in the bSat model the thickness function is related to the saturation scale, this naturally leads to event-by-event saturation scale fluctuations.
	We implement the evolution models in two slightly different forms, given the further constraint that the total normalisation of the proton thickness function has to be conserved. Once a hotspot $i$ is formed, it may either stay unchanged with a fixed width $B_i$ until it splits, or it may continue to become narrower as $B_i=1/|t|$ while keeping its normalisation. The hotspots that are formed in the latter case therefore become hotter with $|t|$ and would thus enhance saturation effect. 
	
	For the bSat model we use a modified profile for hotspots which was proposed in \cite{Kumar:2021zbn}, in order to retain the coherent cross section while introducing fluctuations. 
	
	\section{Results}
	\label{sec:results}
	 \begin{figure*}
	 	\centering
	 	\includegraphics[width=0.35\linewidth]{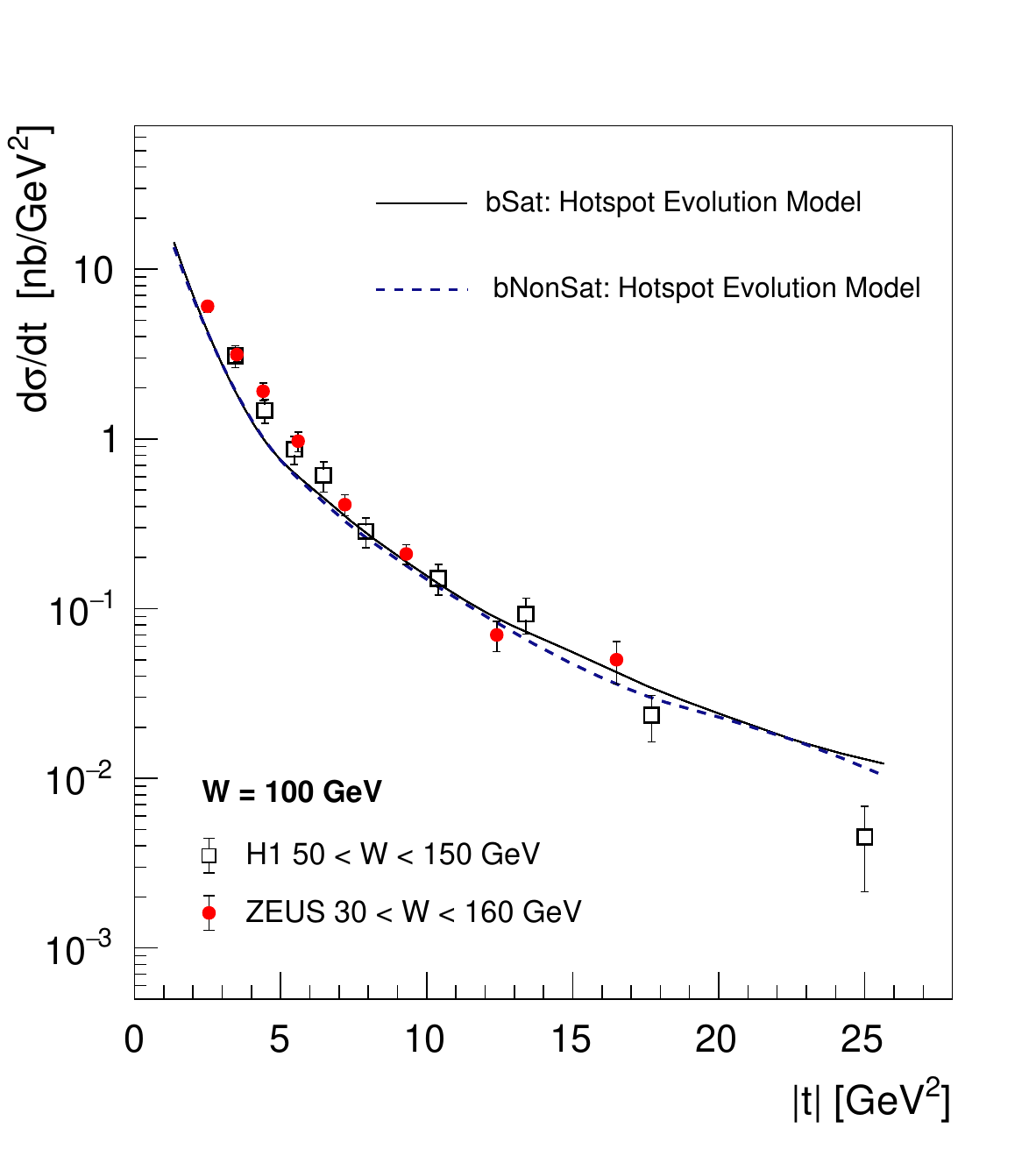} \hskip1.5cm
	 	\includegraphics[width=0.35\linewidth]{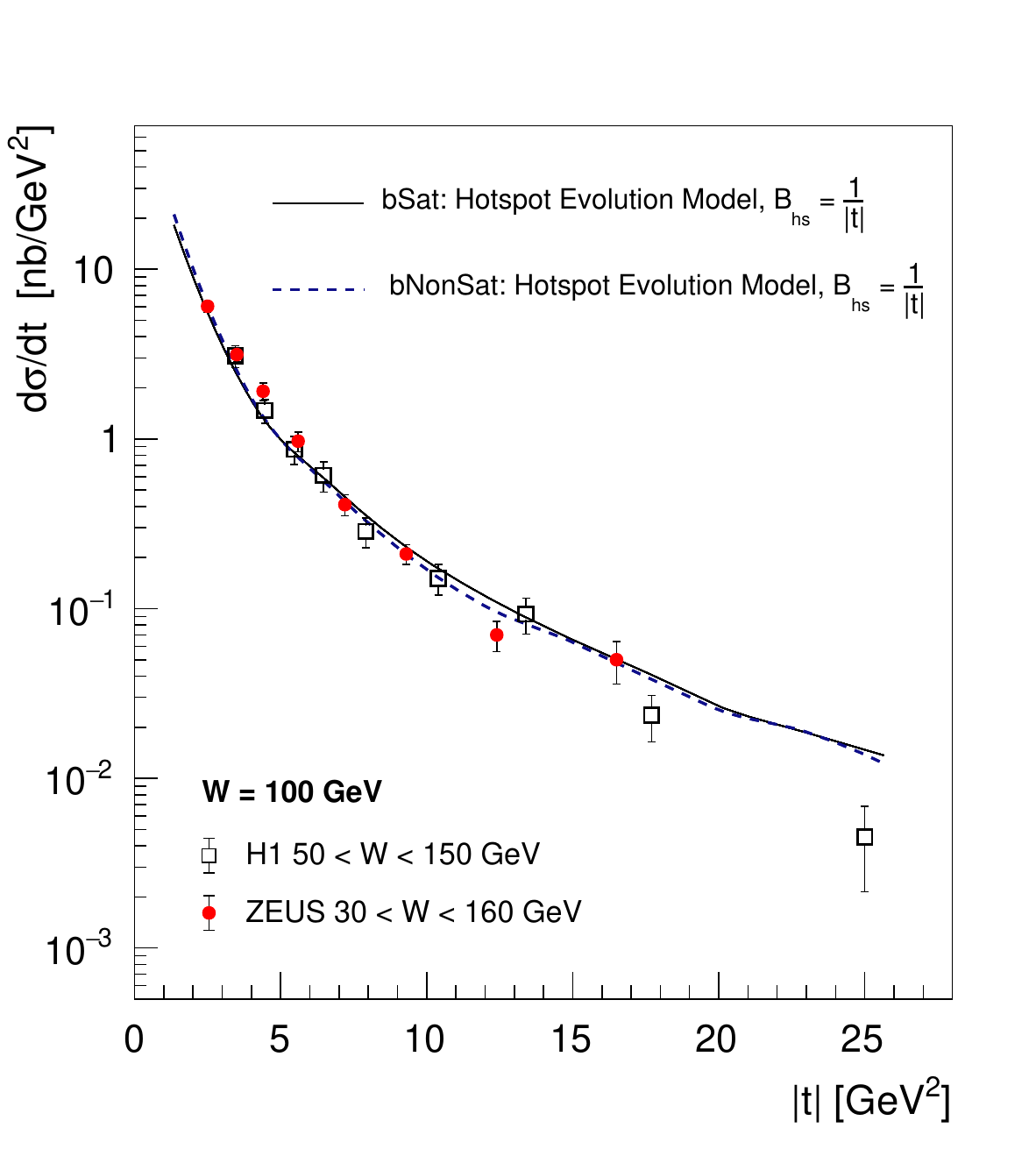}
	 	\caption{The resulting $t$-spectrum from hotspot evolution in the bSat and bNonSat models, compared to measurements from the H1\cite{Aktas:2003zi} and ZEUS \cite{Chekanov:2009ab} experiments. On the left the hotspots remain unchanged between splittings, while on the right they keep evolving with $t$ as described in the text.}
	 	\label{fig:TM}
	 \end{figure*}
	 
	  In Fig. \ref{fig:TM}, we present a comparison of the hotspot evolution model results and the experimental data from HERA for incoherent J/$\psi$ production at large momentum transfers, considering both saturated and non-saturated versions of the dipole model marked bSat and bNonSat respectively. 
	  We checked that the choice of $t_0$ does not affect the results for $t_0\in[0.8, 1.2]~$GeV$^2$. We use $t_0=1.1~\text{GeV}^2$ as the hotspot model  describes the measured data well at this point. 
	  The best fit for the parameter $\alpha$ was found to be $\alpha= 18.5$. This model provides a good description of the measured data. 
	  On the left hand side in Fig.~\ref{fig:TM}, we note that both the saturated and non-saturated versions of the model yield indistinguishable results. In order to allow the hotspots to become hotter, and therefore yield regions with a larger saturation scale, we let the width of all the hotspots become $B=1/|t|$.  
	  The hotspot widths thus evolve even without splitting, but their normalisations remain unchanged. The result of this is shown in the right-hand side of Fig.~\ref{fig:TM}. As can be seen, the description of the data is equally good with this procedure, and there is still no different between bSat and bNonSat. Therefore, we conclude that there is no distinctive non-linear effects present in the $J/\psi$ measurements from HERA. 
	  	  
	 \begin{figure}
		\centering
		\includegraphics[width=0.7\linewidth]{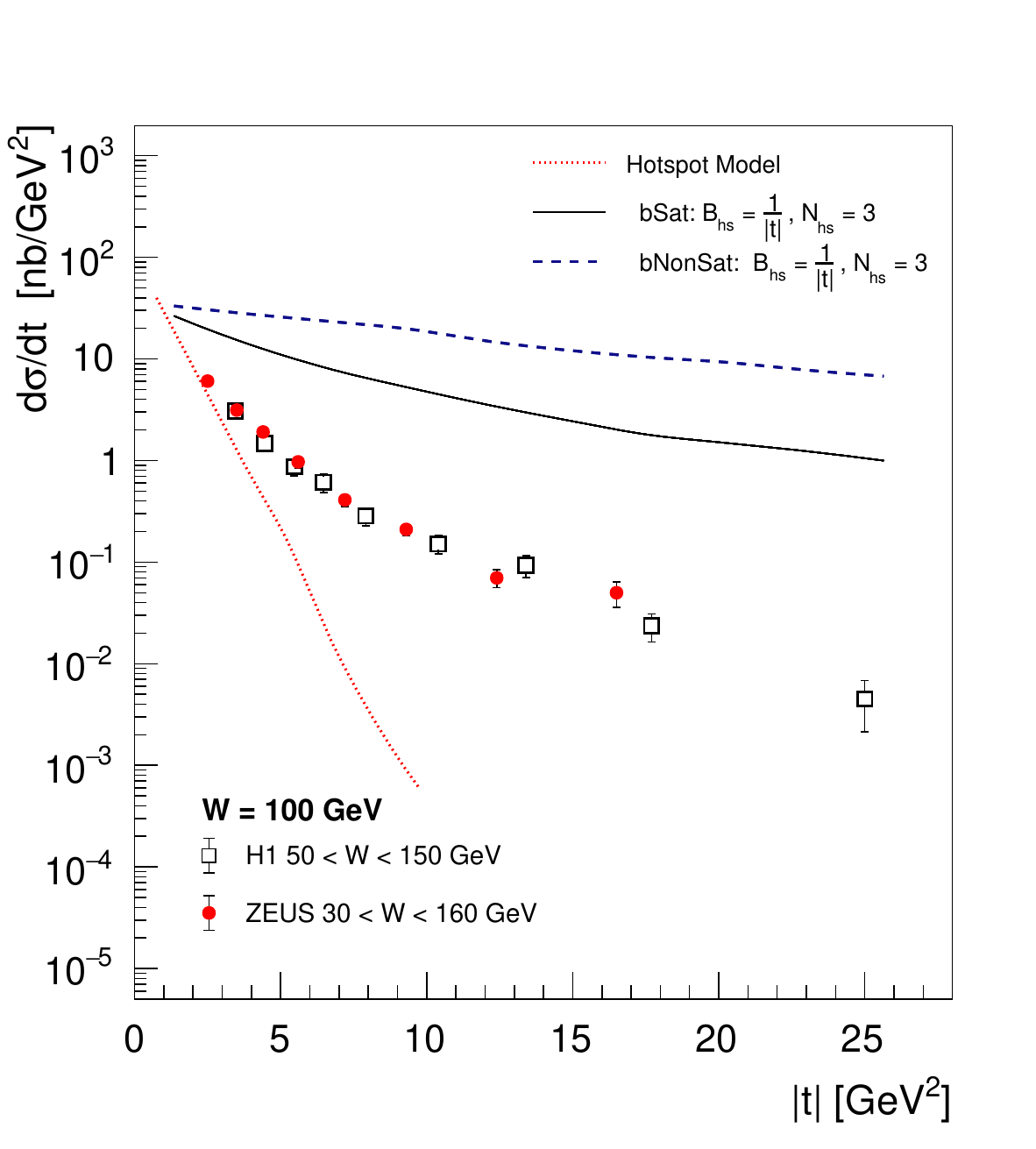} 
		\caption{A comparison between the measured $J/\psi$ $t$-spectrum at HERA with two the hotspot model and with an extreme version of hotspot evolution keeing $N_{hs} = 3$ and letting $B_{hs} = 1/|t|~GeV^{-2}$.}
		\label{model_3}
	\end{figure}
	 In order to better understand the effects from increasing the saturation scale we implement a modified version of the hotspot model in Fig. \ref{model_3}. Here, we let the width of the initial three hotspots decrease as $B_{hs} = 1/|t|~$GeV$^{-2}$, but keep their number constant (no splittings). While the original hotspot model (the dotted red curve) greatly underestimates the HERA data for large $|t|$, its modified version severely overestimates the data for both the bSat and bNonSat models. This is expected, since when the hotspots become narrower as we increase $|t|$, this gives extreme small size fluctuations with very hot hotspots. We can consider this an extreme case with the hottest hotspots possible, while the original hotspot model is the extreme case with the coldest possible hotspots. The HERA data lies inbetween. Here we do see non-linear effects manifest in the $t$-spectrum as there is a clear difference between the bSat and bNonSat models, and this difference is increasing with $|t|$. This corroborates our earlier conclusion that non-linear effects are not present in the measured $t$-spectrum.
\begin{figure*}
\centering
	\includegraphics[width=0.85\linewidth]{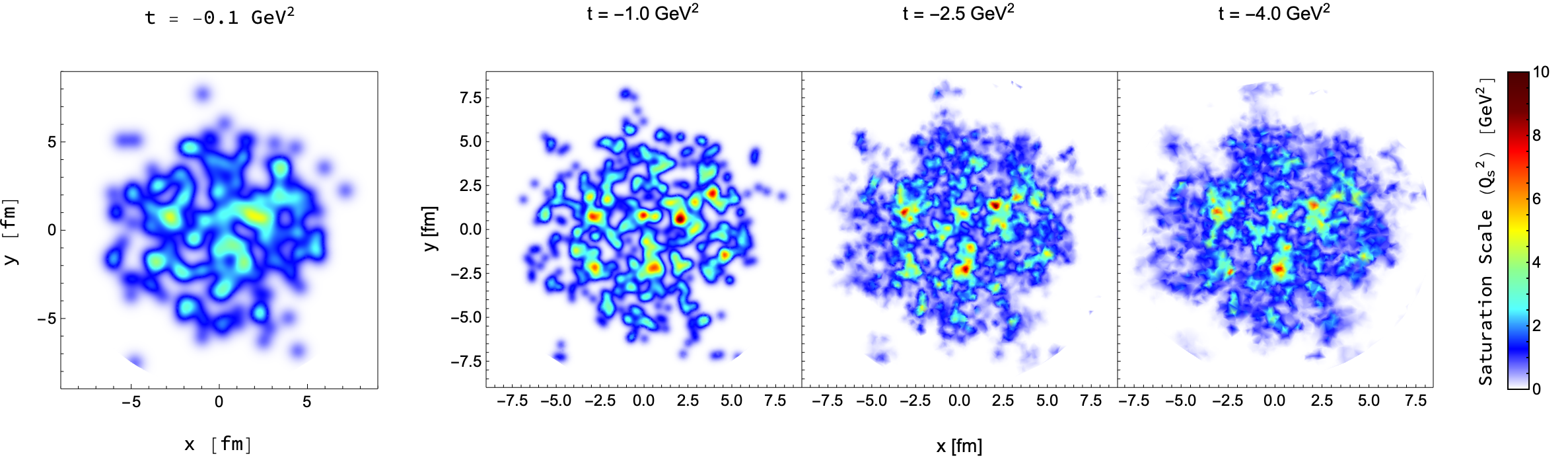} 
	\caption{The resulting saturation scale $Q_S$ as a function of transverse coordinates at different values of $t$ for an initial state nucleon distributions of a lead nucleus. Here, $\xpom=10^{-4}$. The hotspot evolution occurs for $|t|>1.1$~GeV$^2$.}
	\label{fig:QSPb}
\end{figure*}

	 Measuring this incoherent $t$-spectrum at the EIC in $e$Pb collisions as well as in ultra-peripheral PbPb collisions at RHIC and the LHC, may very well yield significant non-linear effects, as we expect an \emph{oomph}-factor in the saturation scale of the nucleus compared to proton's of $Q_S^{(A)}\sim A^{1/3}Q_S^{(p)}$. In order to gauge these effects, in fig. \ref{fig:QSPb}, we show the resulting saturation scale $Q^{(A)}_S(x, \textbf{b})$ from evolving the hotspots in all the nucleons for a initial nucleon configuration of lead. Here we see that we get the hottest regions for moderate values of $|t|$ and that the hotspots tend to diffuse for larger momentum transfers. This gives us an indication of where in the $t$-spectrum one may expect the strongest saturation effects. If we compare the figs. \ref{fig:QSPb} and \ref{fig1} we see a significant oomph-factor in the different saturation scales in protons and heavy nuclei. 
	 
	\section{Conclusions and Outlook}
	We have presented a model for the proton geometry which computes the incoherent cross section for vector meson production at large momentum transfers which is crucial for understanding the fundamental nature of fluctuations in the proton wavefunction, and thereby its spatial structure. The resulting model is consistent with previous observations in the literature, namely that the $ep$ $t$-spectrum can be described with a self-similar gluon wave-function of the proton\cite{Kumar:2021zbn}, and that the small $x$ partons are maximally entangled\cite{Kharzeev:2017qzs,Hentschinski:2023izh}. It has also been observed that extra sources of fluctuations are needed to describe the exclusive diffractive data \cite{Demirci:2022wuy}. Our model naturally introduces several such sources: In the number of hotspots, and in their sizes and normalisations. As the thickness function is directly related to the saturation scale in the bSat model, this naturally leads to event-by-event fluctuations in the saturation scale.
	
	Our model is a classical evolution model, for example similar to a parton shower which solves the DGLAP equations. Likewise, it is a model which is based on the resolution of the measurement, here given by the Mandelstam $t$-variable. One may interpret this as the combined small $x$ gluon wave function collapsing into an area $1/|t|$, revealing whatever structure exist at that resolution. As the resolution is increased, more structure is therefore revealed which manifests as hotspots splitting into smaller hotspots. As a direct consequence from this interpretation, there is a repulsion between hotspots, as they cannot be resolved if they are placed too close together in the transverse plane. A similar repulsion in the hotspot model was introduced ad hoc in \cite{Albacete:2016pmp, Albacete:2017ajt}, in order to explain aspects of flow harmonics seen in high multiplicity $pp$ collisions.

	This hotspot evolution model uses the original hotspot model as a non-perturbative initial state. It then applies a splitting function to determine the probability of the hotspots splitting at a given resolution $t$. This splitting function has one parameter, $\alpha$, which sets the rate of splittings and needs to be fixed by comparison with the measured $t$-spectrum. We start the evolution at a scale $|t_0|=1.1~$GeV$^2$. We emphasise that, apart from the initial state parametrisation which is the hotspot model, we need only one extra parameter to describe the entire $t$-spectrum. The only further assumptions is that of the Gaussian shape of hotspots. 
	It would be desirable as a future project to calculate these splitting functions as well as the hotspot shapes from first principle, which may be possible since the evolution is within the perturbative realm of QCD. Such a first principle evolution model would then be used to restrict the parameters of the initial state hotspot model, such as the number of hotspots, and their widths and shapes.
	
	We noted that there are no discernible difference between the saturated and non-saturated dipole models in their description of the measured data. In order to possibly enhance the difference, we allowed the hotspots to become narrower between splittings such that all hotspots always have the same width $B=1/|t|$ at a given $t$. This allows the longer lived hotspots to become hotter, which locally enhances saturation effects. However, the similarity between the two models remains; we see no non-linear effects manifest in the data. To test the limits of this, we devised a version of the hotspot model designed to maximise saturation scale fluctuations. In this model, the original three hotspots do not split, but only get narrower as $1/|t|$. Here, there is a strong non-linear effect manifesting as the difference between the saturated and non-saturated dipole models, and it grows with $|t|$. However, these models now completely fail at describing the measured data. Therefore we conclude that there are no saturation effects present in the $ep$ $t$-spectrum from HERA. 
	
	This situation is expected to improve in the $eA$ collisions planned at the electron-ion collider (EIC) \cite{Accardi:2012qut, AbdulKhalek:2021gbh}. With it hermetic detectors, the EIC has the potential to measure the large $|t|$ spectrum in $e$Pb collisions. The heavy nucleus' thickness function exhibits increased saturation scale fluctuation due to its geometrical oomph-factor of the saturation scale A$^{1/3}\simeq 6$. We showed that in the hotspot evolution model this will vary even more than in incoherent diffraction. It will be an important future study to extend these results to $e$A collisions at the EIC and perhaps even at the LHeC \cite{LHeC:2020van} and a future Muon Ion Collider \cite{Acosta:2021qpx, Acosta:2022ejc}.
	
		In the bSat model, the thickness function is (almost) directly related to the saturation scale. We have shown that the thickness function depends on both impact parameter and $t$, therefore there is a direct $t$-dependence in the saturation scale as well. We have shown the resulting saturation scales in protons and lead-nuclei as a function of impact parameter at several values of $t$, and the maximal oomph factor are expected to occur around moderate $|t|$ at $\xpom=10^{-4}$. In the future the $\xpom$-, $t$-, and A-dependence of the saturation scale will be studied in detail, which is needed as it has been suggested that the thickness function also has an $\xpom$-dependence \cite{Salazar:2021mpv, Kumar:2022aly}.
	
	\section*{Acknowledgement}
	 This work is supported by the Department of Science \& Technology, India under Grant No. DST/ INSPIRES/03/2018/000344, as well as by the Science and Engineering Research Board Core Research Grant CRG/2022/002507-G.
	
	\bibliographystyle{elsarticle-num}
	\bibliography{mybibfile}
	
	    \end{document}